# All-Optical Control of Magnetization in Quantum-Confined Ultrathin Magnetic Metals

Saeedeh Mokarian Zanjani[1], Muhammad Tahir Naseem[2], Özgür Müstecaplıoğlu[2], Mehmet Cengiz Onbaşlı[1,3*]

[1] Graduate School of Materials Science and Engineering, Koç University, Sarıyer, 34450 Istanbul, Turkey.
[2] Department of Physics, Koç University, Sarıyer, 34450 Istanbul, Turkey.
[3] Department of Electrical and Electronics Engineering, Koç University, Sarıyer, 34450 Istanbul, Turkey.
* Corresponding Author: monbasli@ku.edu.tr

**Abstract:**

**All-optical control dynamics of magnetization in sub-10 nm metallic thin films are investigated, as these films with quantum confinement undergo unique interactions with femtosecond laser pulses. Our theoretical derivations based on the free electron model show that the density of states at Fermi level ($DOS_F$) and electron-phonon coupling coefficients ($G_{ep}$) in ultrathin metals have very high sensitivity to film thickness within a few Angstroms. As $DOS_F$ and $G_{ep}$ depend on thickness, we show that completely different magnetization dynamics characteristics emerge compared with bulk metals. Our model suggests highly-efficient energy transfer from fs laser photons to spin waves due to minimal energy absorption by phonons. This sensitivity to thickness and efficient energy transfer offers an opportunity to obtain ultrafast on-chip magnetization dynamics.**

Quantum confined magnetic nanomaterials such as magnetic ultrathin metals and alloys, and diluted magnetic semiconductors (DMS), provide rich emerging new physics [1-3]. There are also, significant research on quantum confinement effect in the atomic thin semiconductors for novel spin-based photonic quantum technologies and applications [4]. Metallic magnetic thin films have been investigated in applications such as femtosecond laser pulse switching of magnetization [5-8] Elemental magnetic metals with low spin-orbit coupling are ideal for this purpose. The mechanism of all-optical switching (AOS) of magnetization includes the electron bath thermalization after illumination by a fs laser pulse, followed by spin and phonon baths coupling with electrons due to the electron-phonon coupling constant. Magnetic metallic ultrathin films (thicknesses less than 10 nm) behave differently to the fs laser pulse. The electron density of state at Fermi level changes after illumination by an ultrafast single laser pulse due to the quantum confinement [1,9,10]. The free electron theory of metals provides the opportunity to understand the quantum effects associated with the film thickness. Because of its simplicity, the thin-film quantum well is appropriate and provides an introductory justification to the quantum size effects [11]. This effect directly changes the electron heat capacity constant (known as Sommerfeld coefficient, $\gamma$) and coupling between electron and phonon. This concept is defined as all-optical quantum manipulation of magnetization.

In this study, we theoretically investigate the magnetization dynamics for sub-10 nm isolated, free-standing metallic thin films when they are exposed to a femtosecond laser pulse. We assume electron-phonon interactions as shown in figure 1. First, we calculate the electron density of state at Fermi level ($DOS_F$), electron-phonon coupling ($G_{ep}$), and electron heat capacity constant ($\gamma$) and magnetization dynamics in quantum-confined magnetic metals using M3TM (microscopic three temperature model) )

[12,13]. Then, we investigated the variability of magnetization dynamics as functions of film thickness. Previous studies investigated the laser light interaction with magnetic material [14-17]. The laser pulse power is directly transferred to the electron bath, and the electron temperature ($T_e$) increases quickly in the sub-picosecond timescale. Electron thermalization results in a sharp decrease in the magnetization of the thin film. Due to the electron-phonon coupling, $T_e$ balances its energy with a phonon bath, and reaches thermal equilibrium. The magnetization of the film is recovered in the following picoseconds. This energy transfer has been studied with microscopic three temperature model (M3TM) [18,19].

The M3TM does not include coupling of spins with electrons and phonons. The magnetization behavior is captured by the Landau-Lifshitz-Bloch equation (LLB) [20], which describes the time-dependent change of the magnetization after interaction with an fs laser pulse. This model, though, does not study the energy balance between the electron and phonon baths. A more comprehensive model is needed to include both magnetization dynamics and electron and phonon baths equilibrium. The magnetization dynamics is influenced by these energy balance parameters such as $G_{ep}$, $\gamma$, and spin-flip ratio determining the timescales of magnetization change.

The M3TM neglects the spin coupling with electron and phonon. This is advantageous due to eliminating some scattering events such as Elliot-Yafet scattering [21]. In many magnetic materials, weak spin-orbit interactions significantly reduce spin-electron and spin-phonon scattering. Hence, our models here neglect these interactions. Magnetic metallic systems with large spin-orbit coupling, such as transition metal interfaces, 2D electron gas or emergent phenomena such as $SrTiO_3/LaTiO_3$ interfaces which yield emergent superconductivity and large spin-orbit coupling [22], and also transition metal dichalcogenides [23-25] are excluded from the scope of this study.

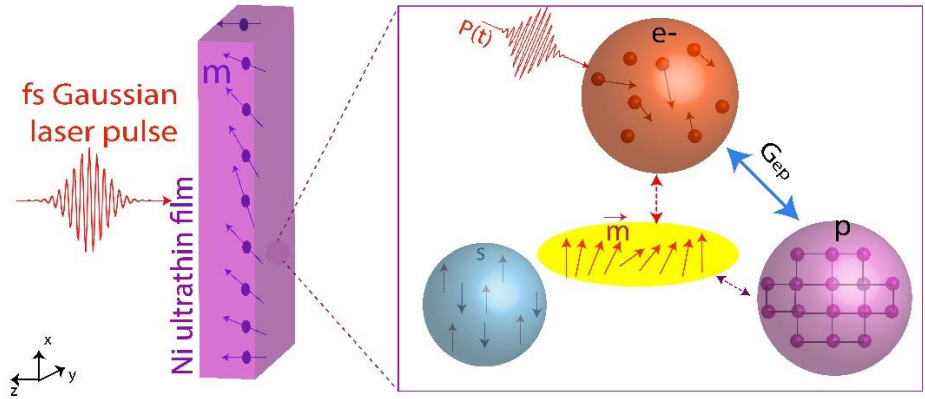

**Figure 1**. The schematic of (Left) ultrashort laser pulse interaction with metallic magnetic ultrathin film. (Right) Coupled interaction of the electron, phonon, and magnetization (extended M3TM).

**Microscopic three temperature model (M3TM) and magnetization dynamics.** We solve the M3TM coupled with magnetization dynamics (extended M3TM) in the Eq. 1-3 based on Koopmans's model [5]. These differential equations describe the energy transfer from femtosecond laser pulse (shown as P(t): time-dependent Gaussian single pulse) to electron, phonon, and magnetization ($m=|M_z|/M_s$).

$$C_e \frac{dT_e}{dt} = -G_{ep}(T_e - T_p) + P(t) \qquad (1)$$

$$C_p \frac{dT_p}{dt} = -G_{ep}(T_p - T_e) \qquad (2)$$

$$\frac{dm}{dt} = Rm \frac{T_p}{T_C} \left(1 - m\coth\left(m\frac{T_C}{T_e}\right)\right) \qquad (3)$$

The parameters used in the Eq. (1)-(3) are shown in table 1, for Ni thin film.

**Table 1.** Parameters used in extended microscopic M3TM

| Parameter | Explanation | Value |
|---|---|---|
| $C_p$ | Heat capacity of phonon | $2.33 \times 10^6$ (J·m$^{-3}$·K$^{-1}$) [5] |
| $C_e$ | Heat capacity of electron (dependent of electron temperature ($T_e$)) | $C_e = \gamma T_e$ (J·m$^{-3}$·K$^{-1}$) [5] |
| $\gamma$ | $\gamma_0 \cdot DOS_F$ | Calculated in part A (J·m$^{-3}$·K$^{-2}$) [1] |
| $\gamma_0$ | Sommerfeld coefficient ($C_p/5T_C$) | 743.22 (J·m$^{-3}$·K$^{-2}$) [5] |
| $G_{ep}$ | $G_0 \cdot DOS_F$ | Thickness-dependent e-p coupling coefficient (See part A) [1] |
| $G_0$ | $(\pi K_B/\hbar) \cdot \lambda(\omega^2)$ | Calculated in part A (W·m$^{-3}$·K$^{-1}$) [1] |
| R | R= Spin-flip ratio = $R_0 \times DOS_F$ | $17.2 \times 10^{12}$ (s$^{-1}$) [5,21] |
| $T_C$ | Curie Temperature of Ni | 627 (K) [5] |

We consider the incoming laser pulse power as a Gaussian single pulse per unit volume as

$$P(t) = \frac{P_0}{\sqrt{2\pi}} \exp\left(-\frac{1}{2}\left(\frac{t}{t_0}\right)^2\right) \quad (4)$$

where $P_0 = \frac{I_0}{d \cdot t_0}$, and $I_0$ and $t_0$ are the laser pulse fluence in J·m$^{-2}$, pulse width (fs), respectively. The injected laser fluence is normalized to a fixed thickness, d, to capture the pulse energy per unit volume. The parameters used in the Eqs. (1)-(3) are shown in table I, for Ni thin film.

To the best of our knowledge, there is no former experimental or modelling study on all-optical control of magnetization dynamics of quantum confined ultrathin metals, which consider the thickness dependence of $G_{ep}$ and $\gamma$ in the extended M3TM. Due to the low thickness of the thin films and quantum confinement effects, the electron-phonon coupling ($G_{ep}$) changes dramatically with the film thickness in the nm regime. Due to the quantum confinement in the electron density of states at the Fermi level ($DOS_F$), both the e-p coupling ($G_{ep}$) and Sommerfeld coefficient ($\gamma$) cannot be considered constant. Our model assumes that $C_p$ is not dependent on the thickness and temperature in the considered time and thickness regime. The spin-flip ratio (R) in Eq. (3) of Ni is a parameter that determines the kinetics of the transient magnetization change. R is dependent of the $G_{ep}$ according to Ref. [5], which is also a function of $DOS_F$ and $L_z$-dependent. In finite-size systems, such as thin metallic films, confinement effects could influence the thermal properties of the lattice phonons, too. Size effects are most significant for thermal transport [26]. On the other hand, thermodynamic properties, such as specific heat, do not change significantly with the size if the temperatures of interest are much lower than the Debye temperature. The primary effect of the finite size is to have fewer states to count in the calculation of specific heat. In strict mathematical terms, the continuum approximation to the states in the k-space cannot be made in finite size systems, and the sum over the states cannot be replaced by an integral. However, the ratio of the exact specific heat determined by the summation to the integral calculation of the specific heat is close to one. For example, it is about ∼ 0.9856 for Aluminum thin film at low temperature, using the Debye model phonon dispersion relation [27]. Accordingly, neglecting the size effects on phonon bath by using the typical bulk value of phonon specific heat is a reasonable approximation in our calculations.

Our model neglects the spin coupling, which leads to the elimination of some scattering events such as Elliot-Yafet scattering [21]. As a result, the fs laser pulse manipulates the magnetization, without excess energy concentration in the lattice, considering the minimal change in the phonon temperature. If we assume that the heat capacity of the nm thick Ni film is equal to the bulk value, with the Gaussian laser pulse fluence of $I_0 = 28$ mJ·m$^{-2}$ and the spot size of 100 μm, the temperature change of a 2 nm-thick Ni film with the density of 8900 Kg·m$^{-3}$, is calculated as $E_{laser} = 2.184 \times 10^{-10}$ J = M × c ×ΔT · M (mass) = $\rho_{Ni}$ ·V = $1.4 \times 10^{-13}$ Kg and the specific heat capacity of Ni (c) is 440 J · (Kg·°C)$^{-1}$. As a result, the temperature change in the thin film will be ΔT = 3.545 °C. The laser fluence ranges for ultrafast magnetization switching of metallic magnetic thin films vary from 1-14 mJ·cm$^{-2}$ (1-14 J·m$^{-2}$) in the literature [18,28,29]. Even if we consider an upper limit of 10 J · m$^{-2}$ as incoming laser energy, the thin film temperature change would be ΔT = 1276 °C, which is still below the melting point of the Ni (1455 °C). Therefore, as long as the fluence is low, we do not cause significant thermal drift in the metal.

## Results

**Effect of film thickness ($L_z$) on chemical potential (μ), Fermi level electronic density of states ($g_F$), electron-phonon, Sommerfeld coefficient (γ), and coupling ($G_{ep}$).** In figure 2(a) and 2(b), the $L_z$-dependence of chemical potential μ (or temperature-dependent Fermi energy) and DOS at Fermi energy are plotted, respectively. Both μ and $g_F$ are dimensionless, normalized with their corresponding bulk values. Figure 2(c) shows $G_{ep}$ changes with $L_z$. We take parameters for Ni for which $\lambda\langle\omega^2\rangle = 49.5$ meV$^2$ and is converted to J$^2$ in our calculations. The depth of the metallic confinement potential is assumed to be $V_z = 10$ eV. The temperature is fixed to $T_e = 5 \times 10^{-3} T_B$, where $T_B = E_B/k_B$ being the Bohr temperature corresponding to the Bohr energy $E_B = 13.6$ eV. Electronic density is used as $n = 3/4\pi r_s^3$ with $r_s = 4a_B$, and $a_B$ being the Bohr radius. The size-dependent oscillations disappear after ∼ 50 Å. This maximum length is a characteristic° value given by $d = 10\lambda_F$, where $\lambda_F$ is the Fermi wavenumber for the bulk. For the electronic density, n, used in the plot, we find d ∼ 50 Å. In addition, the effect of film thickness ($L_z$) on the Sommerfeld coefficient is plotted in figure 2(d). The graphs are plotted for the electronic temperature corresponding to the Bohr energy $E_B = 13.6$ eV. Electronic density is used as $n = 3/4\pi r_s^3$ with $r_s = 4a_B$, and $a_B$ being the Bohr radius. Similar plots are included in the Supplementary Information for the electron temperature of $T_e = 0$ K. The depth of the metallic confinement potential is taken to be $V_z = 10$ eV. The $V_z$ value is theoretically infinite, however, in the FEM simulations, its value is considered as a finite number comparable to its bulk Fermi value [10]. In the following sections, we have shown the results of our calculations of μ, $g_F$, γ, $G_{ep}$, as well as magnetization dynamics, based on free electron model (FEM), for $V_z$=10 eV. The results of our calculations for different potential well depths ($V_z$ =5, 15, and 20 eV) are also included in Supplementary Information.

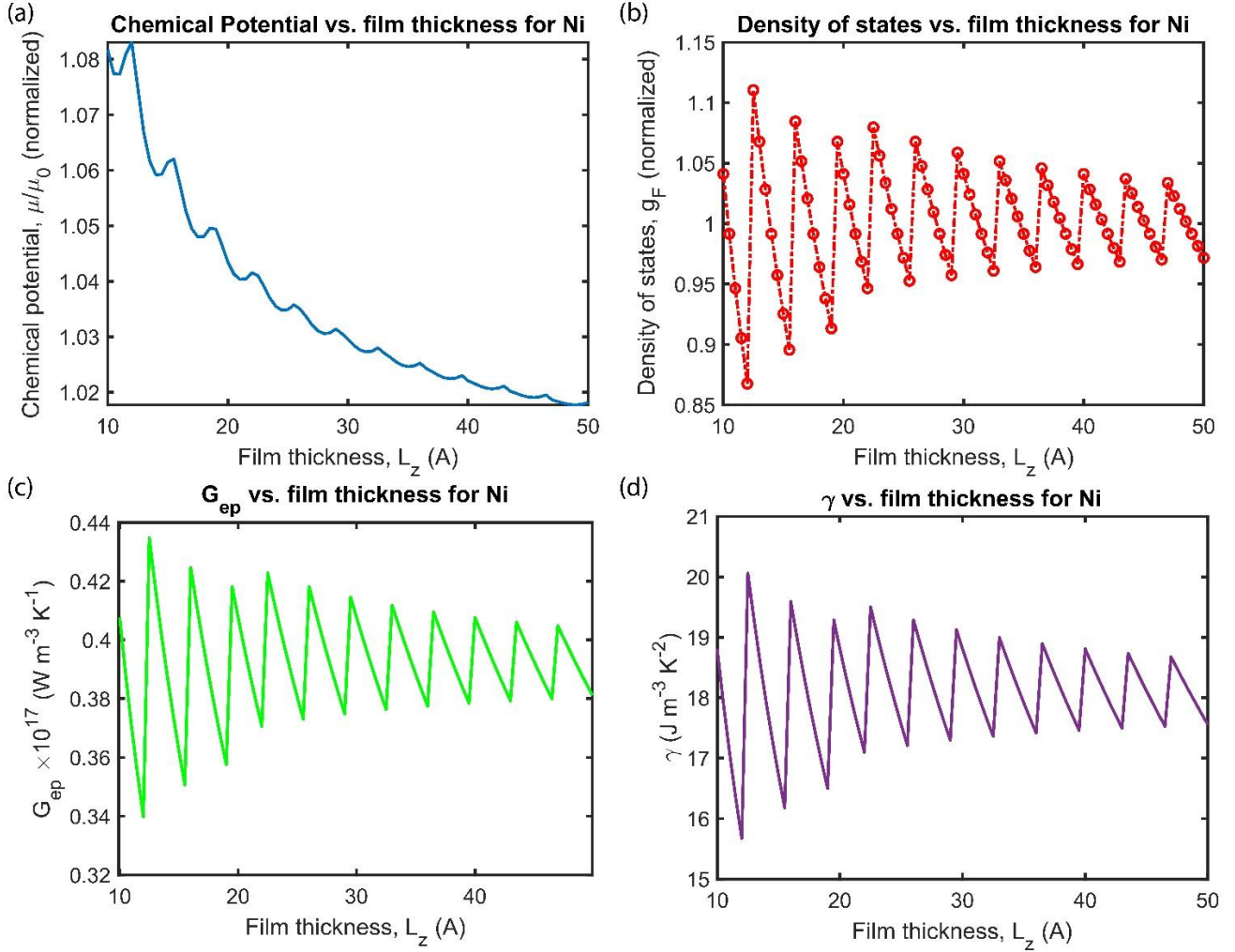

**Figure 2. Thickness dependent quantum-confinement effect.** (**a**) Chemical potential (finite temperature Fermi energy) and (**b**) Density of states (DOS) at the Fermi energy as a function of ultrathin film thickness $L_z$. Both m and $g_F$ are dimensionless, normalized with their corresponding bulk values. (**c**) Ultrathin film thickness-dependence of $G_{ep}$, electron-phonon coupling coefficient and (**d**) $\gamma$ (Sommerfeld coefficient) as a function of $L_z$.

Free electron model (FEM) perfectly reflects the oscillations in the $DOS_F$, $G_{ep}$, and $C_e$ as the result of quantum confinement for the metals whose valance electrons lie in p-band such as Al. Since Ni is a transition metal with an almost full d-band structure[1], the possible complexity in configuration of electrons in the Fermi level would require beyond free electron methods to determine the DOS, such as density functional theory (DFT). However, FEM still captures the essential physics behind the quantum size effects even for transition metals and more complex structures [11]. Ab initio methods confirm the conclusions of the free electron model [30,31].

The oscillations of $G_{ep}$ and $\gamma$ are due to the Fermi level oscillations translated to DOS, arising from the quantum confinement (cf. Eq. 4 in the methods section). This essential physics (discreteness of the $k_z$) remains the same irrespective of the simplicity or complexity of the Fermi surface. Failure of the continuum approximation in the confined direction yields discrete plateaus for the electronic states. Accordingly, the formation of such quantum well states (QWS) in the confined direction captures the basic physics of the quantum size effects, manifested as size-dependent oscillations in physical observables[32-35]. We show the Fermi level oscillations in figure 2(c), which translates to $DOS_F$, hence to $G_{ep}$ and $C_e$. For Ni, although the d-band structure leads to a more complicated Fermi level than Al, this effect decreases the magnitude of $G_{ep}$ compared to the values used in the experimental studies [1]. Even if the decrease in these terms are not entirely in agreement quantitatively with the reported values, our model still reflects the effect qualitatively. FEM predicts quantum size effect oscillations in the magnetization but should be regarded as a qualitative description for the magnetic metals with more complex Fermi configurations such as Ni. The effect can be studied more rigorously and quantitatively using ab initio calculations of the band structure, Fermi level, and DOS. Furthermore, it can be optimized by considering more complex materials using beyond FEM analysis. In this work, we will be presenting the essential physics and predicting quantum size effect in magnetization in the same spirit of quantum size effects in electronic conduction.

**Microscopic three temperature model coupled with magnetization dynamics and spectral response.** In figure 3, the magnetization dynamics and transient electron and phonon temperatures ($T_e$, and $T_p$, respectively) resulting from the extended M3TM are shown for 20 Å Ni thin film. According to figure 3, illuminated with a Gaussian single laser pulse of $I_0 = 28$ mJ·m$^{-2}$, the magnetization of the Ni thin film decreases in sub 100 fs due to the thermalization of the electron bath and E-Y scattering. Due to the low heat capacity of the electrons, $T_e$ reaches $1.5T_c$ of Ni (940.5 K). However, due to electron-phonon coupling, $T_e$ cools down to an equilibrium temperature with phonon (lattice) in 200 fs, and the magnetization recovers to close to its initial value (> 96%), in around 1 ps.

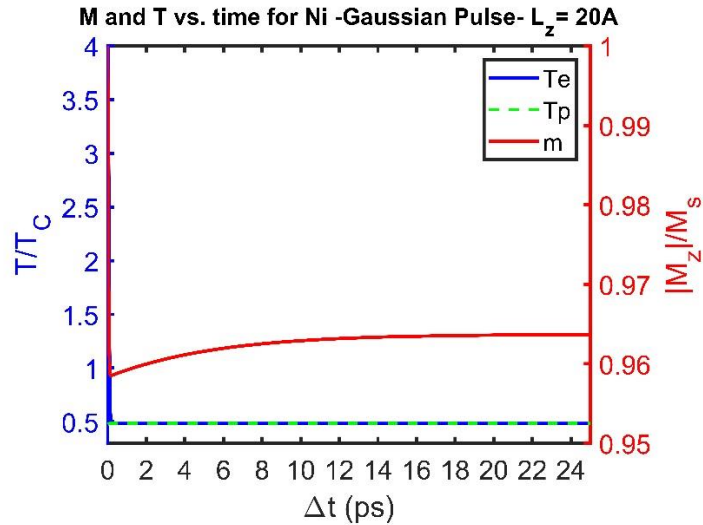

**Figure 3**. Transient $T_e$, $T_p$, and normalized magnetization (m = $|M_z|/M_s$) for 20 Å thick Ni film illuminated with $I_0 = 28$ mJ·m$^{-2}$ Gaussian single laser pulse.

**Dependence of the magnetization dynamics on film thickness ($L_z$).** Figure 4(a) shows the thickness dependence of magnetization dynamics in the extended M3TM for $L_z$ = 10 Å, 15 Å, 20 Å, 25 Å, 30 Å, 35 Å, 40 Å, 45 Å, and 50 Å thick Ni films. According to this figure, the film thickness has a minimal influence on timescales of the demagnetization and recovery. However, it changes the demagnetization ratio (the dip on the magnetization curve) dramatically. The effect of thickness on the magnetization dip is shown in figure 4(b) for different $L_z$ values changing from 10−50 Å (with a step size of 2 Å) figure 2, due to the dramatic change of the $DOS_F$ at sub-5nm (50 Å) thickness regime, the electron- phonon coupling, and Sommerfeld coefficient change considerably with increasing film thickness. The dips have the same position and only their intensities change. This indicates that the change in the rate ($G_{ep}$) triggers a stronger magnetization loss/recovery without altering the spin wave emission spectra.

Figure 4(c) and 4(d) show the dependence of $T_e$ and $T_p$ maxima on the film thickness. The maximum phonon temperature does not change dramatically with the film thickness; however, the modulation of the $T_e$ maxima with increasing the film thickness is similar to the behavior of $G_{ep}$ and demagnetization dip. The electron temperature and the magnetization dip change dramatically from 22 to 22.5 Å (or 25.5 to 26 Å, 29 to 29.5 Å, 32.5 to 33 Å, 36 to 36.6 Å, 39.5 to 40 Å, 43 to 43.5 Å, 46.5 to 46 Å). The electron temperature can go up to 3300 K depending on $L_z$. The extreme sensitivity of electron temperature (i.e. 3300 to 2620 K between) to the thickness $L_z$ (22 to 22.5 Å) shows that piezoelectric modulation can be a viable method for controlling "hot electrons". The fact that electron temperature exceeds the Curie temperature does not prevent the nanomagnets from recovering magnetization. Due to increasing electron temperature, chemical potential gradient causes charge currents within metallic nanomagnets (see μ as a function of $T_e$ in the Supplementary figure S2 (a)). Still, since the highest electron temperature never exceeds 3300 K, the chemical potential difference is less than 1%.

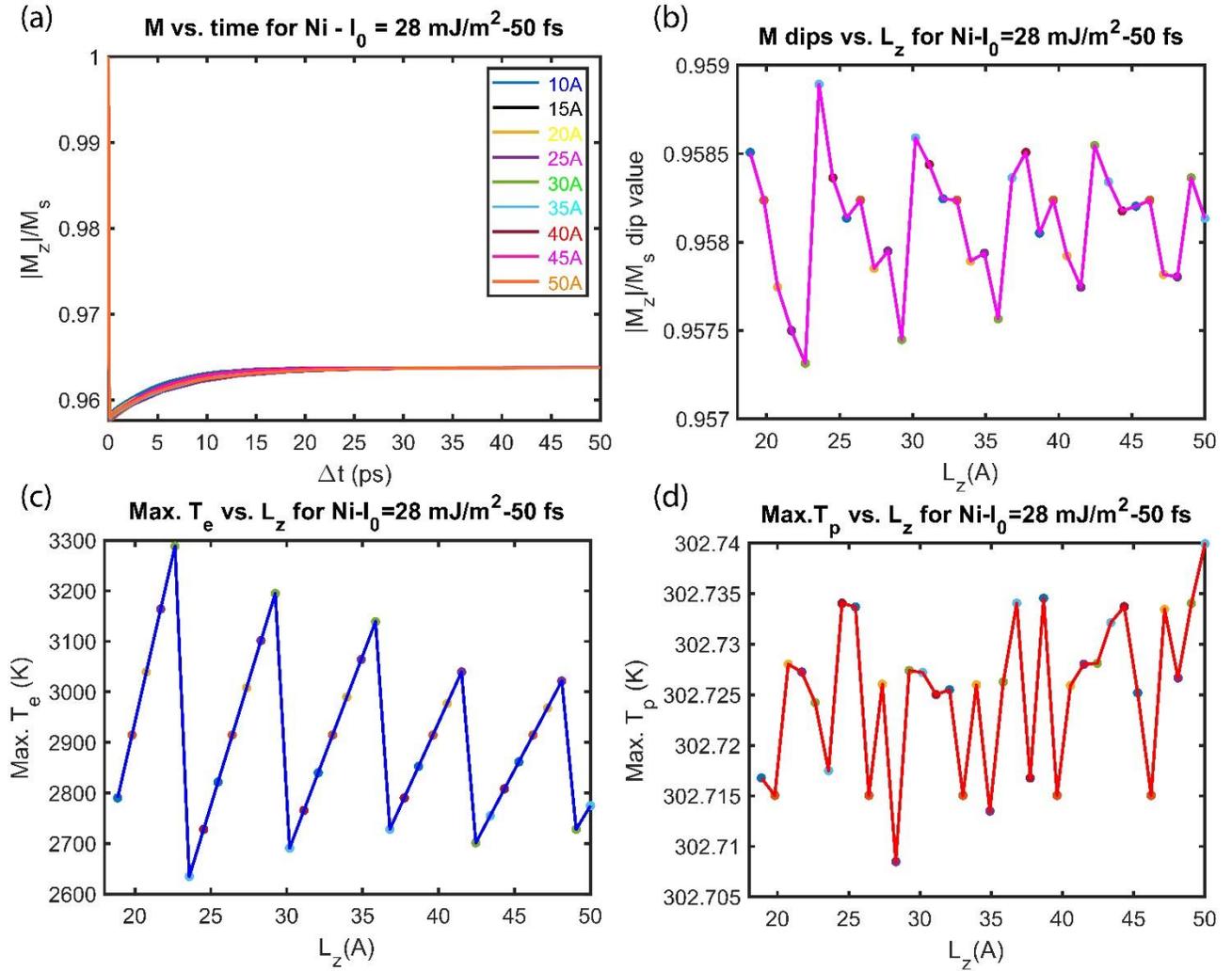

**Figure 4.** Effect of Ni film thickness on (**a**) magnetization dynamics for $L_z$ = 10 Å, 15 Å, 20 Å, 25 Å, 30 Å, 35 Å, 40 Å, 45 Å, and 50 Å thick Ni films, (**b**) demagnetization dip, (**c**) $T_e$, and (**d**) $T_p$, illuminated

with $I_0 = 28$ mJ·m$^{-2}$ Gaussian single laser pulse. The inset of the figure 4 (**a**) is included in Supplementary Information (Figure S3).

As presented in the Supplementary Information (figure S4-S7), the finite size of the potential well, does not change the qualitative predictions of our model. The calculations results for the various potential well depths ($V_z$=5, 10, 15, and 20 eV) shows that only in case of sub-20 Å film thicknesses, $V_z$ size might result in a few percent difference in the calculations. However, it does not influence the qualitative predictions of our model on decrease and oscillations in the electron density of states and electron-phonon coupling as the result of quantum confinement effect. In other words μ increases slightly (~ 3%), in sub-20 Å range, from $V_z$=5 eV to $V_z$=10 eV, but the oscillations are not considerable when $V_z$ changes from 10 eV to 15 eV or 20 eV. Accordingly, $g_F$, $G_{ep}$, and γ oscillations decrease (~ 4%) changing from $V_z$=5 eV to $V_z$=10 eV, while remains unchanged for other $V_z$ values. We also investigated the sensitivity of the figure 4 results to the size of $V_z$. Despite similar negligible variations for $V_z$= 5 eV in case of low film thicknesses, changing the depth of the potential well did not influence the conclusions of the quantum confinement effect on the oscillations of the magnetization dynamics as well as electron and phonon temperatures (see Supplementary information figure S6-S7). Since the decrease (inversely proportional to the film thickness) in the chemical potential is averagely monotonic [10] (the amplitude of the oscillations does not exceed a few percent), the qualitative results of our calculations are consistent using different potential well depths. In other words, independent of the quantum well depth, quantum confinement effect reduces $G_{ep}$, and consequently effects the transient magnetization behavior as well as the electron and phonon temperatures, as discussed extensively in the previous part.

The quantization of energy levels in the direction perpendicular to the film due to the thin film's electron confinement leads to the oscillatory dependence of many properties on the film thickness due to quantum size effects. Despite the quantitative small oscillations as the result of quantum confinement, the effect is measurable by various experimental means such as magneto-thermoelectric measurements [36], Kelvin probe force microscopy [37], Hall-bar magnetoresistance measurement [38], and scanning tunneling spectroscopy (STS) technique [39]. These methods are reported to measure the quantum size effect in different metallic and also more complex thin films.

**Energy transfer from fs laser pulse to magnon system.** Since spin-orbit and spin-phonon couplings are neglected, all the laser pulse energy goes to the phonon bath after equilibration with the electron bath. So the energy absorbed by the phonon is equal to $E_p = C_p \cdot \Delta T_p$. The maximum phonon temperature change is for the film with the thickness around 35 Å, which is around° 0.022 K. So the energy absorbed by the phonons is simply $E_p = (2.33 \times 10^6$ J·m$^{-3}$·K$) \times (\pi/4 \times (100 \times 10^{-6}$ m$)^2) \times (3.5 \times 10^{-9}$ m$) \times (0.022$ K$)$ $=1.4 \times 10^{-12}$ J. From thermodynamic standpoint, phonon energy change (heat) is the difference between the injected energy into the system from laser pulse and the energy change in the spins (work done). The efficiency of work done by laser pulse on spins is $\eta = 1 - Q_{phonon}/Q_{laser}$. The absorbed energy for increasing the lattice temperature is $Q_{phonon} =1.4 \times 10^{-12}$ J and the laser energy is $2.184 \times 10^{-10}$ J. The efficiency is $\eta = 99.36\%$. In Supplementary Information, we considered a condition where the spin-electron, and spin-phonon scattering are not neglected. Since a part of a laser energy is lost due to the scattering phenomena, the laser energy needed to manipulate and recover the magnetization increases. Considering the laser fluence of 35 mJ·m$^{-2}$ ($Q_{laser} = 2.73 \times 10^{-10}$ J) the increase in the phonon temperature maxima for 35A° film does not considerably change, according to the figure S8 (b) in the Supplementary Information. However, due to the excess laser energy needed for recovery of the magnetization, the work done by the laser on the spins increases which increases the efficiency to $\eta = 99.48\%$.

**Conclusion**

We investigated the all-optical magnetization dynamics in the quantum confined magnetic Ni ultrathin films. Our theoretical model shows that due to the quantum confinement in the films with the thicknesses of a few tens of Angstrom, the electron-phonon coupling coefficient ($G_{ep}$) in the M3TM is highly sensitive to the film thickness and could not be considered constant. This effect changes the amount of magnetization drop after interaction with the fs laser pulse, but not the timescales of the magnetization. In addition, we show that our qualitative predictions of magnetization dynamics in the quantum-confined ultrathin magnetic films are not sensitive to the size of the quantum well.

We show that, laser-induced magnetization dynamics could drive ultrafast exchange-driven magnetization oscillations [40]. In addition, the quantum confinement effect decreases the lattice temperature change due to the lower laser fluences needed for magnetization control in Ni ultrathin film. Thus, energy efficiency of exciting spin waves with lasers could be enhanced [41].

Our study shows that the choice of the film thickness in the Angstrom regime could help modulation of magnetization with around three orders of magnitude lower laser fluences compared to the reported experimental values. We also show that the energy transfer rate from the laser pulse to the lattice is so low that the lattice temperature stays far below both Curie temperature and the melting point of the Ni.

**Methods**

**Quantum confinement effects on DOS, $G_{ep}$, and γ.** Our physical system, which consists of a ferromagnetic ultrathin metal illuminated by an ultrashort (femtosecond) laser pulse, can be microscopically described by a generic Hamiltonian for interacting electron and phonon baths that reads,

$$H = H_e + H_p + H_{ep} \quad (1)$$

where $H_e$ and $H_p$ are the Hamiltonians of free electrons and phonons, respectively. $H_{ep}$ models the electron-scattering from the lattice, ignoring the spin flips. We remark that the baths are, in fact, not exactly free. The calculation of the electron-phonon coupling parameter $G_{ep}$ in the two-temperature model proceeds by application of the Fermi's golden rule using $H_{ep}$. In the Supplementary Information, we present a more extensive review on the derivation in Ref. [9], which is based upon pioneering works [42]. It allows for including beyond free electron theory effects and arbitrary DOS.

We can write the temperature gradient between the baths such that

$$C_e \frac{dT_e}{dt} = -G_{ep}(T_e - T_p) \quad (2)$$

Electron-phonon coupling factor $G_{ep}$ is given by

$$G_{ep} = \pi \hbar \lambda \langle \omega^2 \rangle g_F \quad (3)$$

where $k_B$ is the Boltzmann constant, λ is the electron-phonon mass enhancement parameter [43], and $\langle \omega^2 \rangle$ is the second moment of the phonon [44]. At low temperatures, we can take $C_e = \gamma T_e$, where $\gamma = \pi^2 k_B^2 g_F / 3$, according to the Sommerfeld expansion [45]. Numerical examination of $C_e$ for different metals at higher temperatures, which is beyond the scope of the present contribution, can be found in the literature [1]. A similar equation can be obtained for the phonon bath. These equations can be changed to temperature rate equations using the corresponding specific heats. Finally, introducing the additional terms for the laser pulse absorption and heat diffusion, the two-temperature model [46] can be developed.

The sensitivity of $g_F$ to the variations of the thickness of the ultrathin film, $L_z$, allows for additional control over the electron-phonon coupling $G_{ep}$. We use the free electron model, while the electrons are confined in a potential well of size $L_z$ and depth $V_0$. This limits $k_z$ to be always less than $k_{top} = (2mV_0)^{1/2}/\hbar$.

Quantization of $k_z$, according to [10], reduces the finding temperature-dependent Fermi energy is a counting problem of electrons living on disks confined in the Fermi sphere with temperature-dependent radius.

$$K_z L_z = n_z \pi - 2 \sin^{-1}\left(\frac{K_z}{K_{top}}\right) \quad (4)$$

Allowed $k_z$ form a set $k_z$ whose dimension is temperature dependent. The physical reason for the oscillations with $L_z$ is due to the discreteness of the electronic wave number along the finite thickness direction, $k_z$.

At zero temperature, the number of electrons can be determined by [10].

$$N = 2\frac{L_x L_y}{4\pi^2}\sum_{k_z \leq k_F} \pi(k_F^2 - k_z^2) \quad (5)$$

where $L_x$, $L_y$ are the long, transverse sizes of the ultrathin film ($L_z \ll L_x, L_y$). At finite temperatures, we should use $N = 2\sum_k f_k$, where the Fermi-Dirac distribution would contain temperature-dependent Fermi energy $\mu(T)$ such that

$$N = 2\frac{L_x L_y}{4\pi^2}\sum_{k_z \epsilon K_z} \int dk_x dk_y \frac{1}{e^{(E(k)-\mu(T))/k_B T}+1} \quad (6)$$

Using

$$E(k) = \frac{\hbar^2}{2m}(k_x^2 + k_y^2 + k_z^2) \quad (7)$$

the integral can be evaluated analytically such that

$$n = \frac{T}{\pi L_z}\sum_{k_z \epsilon K_z} \ln\left[1 + \exp\left(\frac{k_\mu^2 - k_z^2}{T}\right)\right] \quad (8)$$

where $k_\mu$ is the temperature-dependent Fermi wavenumber. We use scaled variables such that the length scale is $a_B$, the temperature scale is $T_B$, and the energy scale is $E_B$. This equation generalizes the zero temperature counting problem in Eq. 5, which determines the quantum confinement effect on DOS at Fermi energy to finite temperatures. In Supplementary Information, we describe a method in Ref. [10] with which one could find how the $L_z$ dependence of DOS at Fermi level changes with temperature, to be able to take into account this size effect for $G_{ep}$. At the range of temperatures we are interested in, the temperature dependence of $G_{ep}$ is negligibly different than the zero temperature results.

**Acknowledgments**

The authors a gratefully acknowledge Dr. Turan Birol for his valuable consultations. Funding support from European Research Council Starting project SKYNOLIMIT, Grant No. 948063 and TUBITAK grant no. 120F230 and TUBITAK Grant No. 117F416 is gratefully acknowledged.

**Author Contributions**

M.C.O. and Ö.M. designed the study. S.M.Z. performed the modeling, and analyzed the results with help from M.C.O and Ö.M. Ö.M. performed the modeling of µ, $DOS_F$ and $G_{ep}$ based on the free electron model in figure 2. S.M.Z., Ö.M., and M.C.O. wrote the manuscript. All authors discussed the results and commented on the manuscript.

**Competing Interests**

The authors declare no competing interests.

# Supplementary Information for "All-Optical Control of Magnetization in Quantum-Confined Ultrathin Magnetic Metals"

Saeedeh Mokarian Zanjani[1], Muhammad Tahir Naseem[2], Özgür Müstecaplıoğlu[2], Mehmet Cengiz Onbaşlı[1,3*]

[1]Graduate School of Materials Science and Engineering, Koç University, Sarıyer, 34450 Istanbul, Turkey.
[2] Department of Physics, Koç University, Sarıyer, 34450 Istanbul, Turkey.
[3] Department of Electrical and Electronics Engineering, Koç University, Sarıyer, 34450 Istanbul, Turkey.
[*] Corresponding Author: monbasli@ku.edu.tr

This supplementary information is designed in three main parts. In the first part, we present the complete derivation of the three temperature model, which we explained and used in the main manuscript. In the second part, we compare the thickness dependence of the chemical potential and the Fermi level electronic density of states for two different electron temperature ($T_e$). Furthermore, we investigate the sensitivity of our calculation results presented in the main paper to the potential well depth ($V_z$). In the third part, we present the condition where we involve the spin-electron and spin-phonon coupling to justify the ability of our model to explain the high efficiency of energy transfer to the spin system using quantum-confinement in the all-optical magnetization manipulation.

**Derivation of the microscopic three temperature model**

Here we present a brief derivation of the microscopic three temperature model based on the pioneering works [1,2], including free electron theory. Then the effect of ultrathin film thickness on the electron density of state at the Fermi level, and eventually, the electron-phonon coupling coefficient is theoretically modeled.

This section is organized to present our theoretical derivation of the microscopic three temperature model. A ferromagnetic ultrathin metal illuminated by an ultrashort (femtosecond) laser pulse can be microscopically described by a generic Hamiltonian for interacting electron and phonon baths as the following:

$$H = H_e + H_p + H_{ep} \qquad (S1)$$

where $H_e$ and $H_p$ are the Hamiltonians of free electrons and phonons, respectively. $H_{ep}$ models the electron-scattering from the lattice, ignoring the spin flips. We remark that the baths are, in fact, not exactly free. The interactions in the same subsystems $H_{ee}$ and $H_{pp}$ lead to fast equilibration and hence, they are dropped after instant thermalization assumptions. Their effects are implicit in the dynamics; we will not express them here explicitly. Ignoring the spin degree of freedom of the electrons, non-interacting gas of them has the Hamiltonian

$$H_e = \sum_k E(\mathbf{k}) c_\mathbf{k}^\dagger c_\mathbf{k} \qquad (S2)$$

Here $c_\mathbf{k}$ ($c^\dagger_\mathbf{k}$) annihilates (generates) an electron with excitation energy $E(\mathbf{k})$ in Bloch state $|\mathbf{k}\rangle$. Initially, the gas in metal is in equilibrium at an ambient temperature $T_0$. After the ultrashort pulse excitation, this gas can quickly reach a new temperature due to (screened) Coulomb interaction $H_{ee}$.

An ensemble of non-interacting quantum harmonic oscillators describes the bath of lattice vibrations, phonons, with the Hamiltonian

$$H_p = \sum_q \hbar\omega_q (a_q^\dagger a_q + 1/2) \qquad (S3)$$

Here the operators $a^\dagger_q$ and $a_q$ generate and annihilate phonons with quasi-momentum $\mathbf{q}$, respectively. Phonons obey the Bose-Einstein statistics. Einstein model is the simplest choice to describe them, while Debye model is also possible.

Fröchlich interaction describes the scattering of spinless electrons from the lattice as

$$H_{ep} = \sum_{kk'q} V_{kk'q} c_k^\dagger c_{k'} (a_q^\dagger + a_q) \qquad (S4)$$

where we denote the matrix element of the scattering process with $V_{kk'q}$.

The calculation of electron-phonon coupling parameter $G_{ep}$ in the two-temperature model proceeds applying of the Fermi's golden rule using $H_{ep}$. We will present a brief review of the derivation in Ref. [1], which is based upon pioneering works [2], and allows for including beyond free electron theory effects and arbitrary DOS. Accordingly, we write the energy transfer rate from electrons to the lattice $E_e$ as

$$\frac{\partial E_e}{\partial t} = \frac{4\pi}{\hbar}\sum_{kk'}\hbar\omega_q |V_{kk'}|^2 S(k,k')\delta(E(k) - E(k') + \hbar\omega_q) \qquad (S5)$$

Here,

$$S(k,k') = (f_k - f_{k'})n_q - f_{k'}(1 - f_k) \qquad (S6)$$

is the thermal factor, that depends on the Fermi-Dirac ($f_k$) and Bose-Einstein ($n_q$) distribution functions of the electron and phonon baths in their "local" thermal equilibrium states [1]. Explicitly they are defined as $f_k = 1/(1 + \exp(E(k) - \mu)/k_B T_e)$ and $n_q = 1/(\exp(\hbar\omega_q/k_B T_p) - 1)$. Near room temperature, after the conversion of the k summation to continuum energy integrals, we can write

$$\frac{\partial E_e}{\partial t} = 2\pi g_F \int d\Omega \alpha^2 F(\Omega)(\hbar\Omega)^2 [n(\hbar\Omega, T_p) - n(\hbar\Omega, T_e)] \qquad (S7)$$

with $g_F$ denoting the electronic DOS at Fermi level. The limitation of the DOS to the Fermi level value assumes that only those electrons near the Fermi energy could contribute to the scattering process from the lattice vibrations around room temperature. The term $\alpha^2 F(\Omega)$ is the Eliashberg spectral function [3]. If we assume temperatures are below Debye temperatures but higher than phonon mode energy $\hbar\Omega \ll k_B T_p, k_B T_e$, then we can replace the population distributions with a temperature gradient between the baths such that

$$C_e \frac{dT_e}{dt} = -G_{ep}(T_e - T_p) \qquad (S8)$$

where we used $dE_e = C_e dT_e$. Electron-phonon coupling factor $G_{ep}$ is given by

$$G_{ep} = \pi\hbar\lambda\langle\omega^2\rangle g_F \qquad (S9)$$

where $k_B$ is the Boltzmann constant, $\lambda$ is the electron-phonon mass enhancement parameter [4], and $\langle\omega^2\rangle$ is the second moment of the phonon [3]. At low temperatures, we can take $C_e = \gamma T_e$, where $\gamma = \pi^2 k_B^2 g_F/3$, according to the Sommerfeld expansion [5]. Numerical examination of $C_e$ for different metals at higher temperatures, which is beyond the scope of the present contribution, can be found in the literature [6].

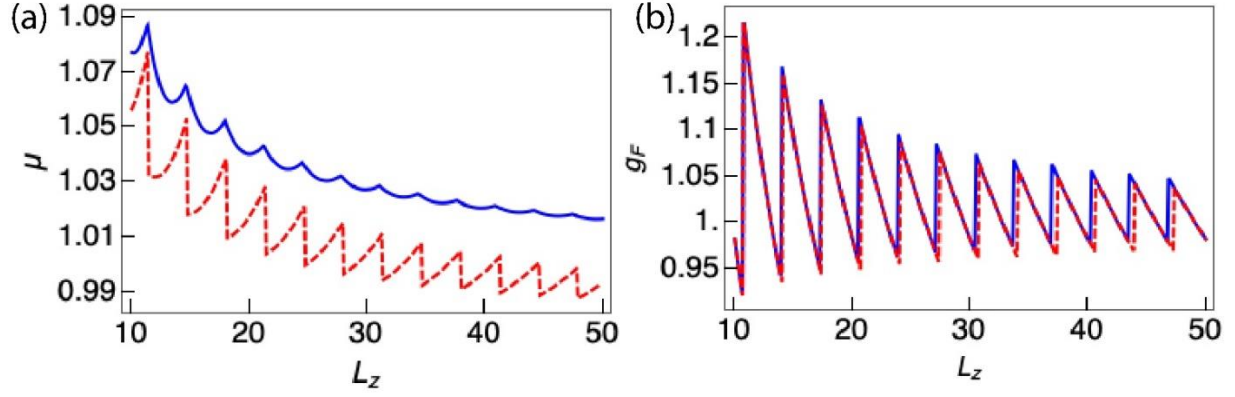

**Figure S1**. (**a**) Chemical potential (finite temperature Fermi energy) μ and (**b**) Density of states (DOS) at the Fermi energy as a function of ultrathinfilm thickness Lz (in units of A). Both μ and $g_F$ are dimensionless, normalized with their corresponding bulk values. Solid blue lines are the zero temperature results. Dashed red lines are for the electronic temperature $T_e = 5 \times 10^{-3} T_B$, where $T_B = E_B/k_B$ being the Bohr temperature corresponding to the Bohr energy $E_B = 13.6$ eV. Depth of the metallic confinement potential is taken to be $V_z = 10$ eV. Electronic density is used as $n = 3/4\pi r_s^3$ with $r_s = 4a_B$, and $a_B$ being the Bohr radius.

### Dependence of DOS$_F$ on temperature

In this section, we describe a theoretical method with which the dependence of DOS at Fermi level changes with temperature, to be able to take into account this size effect for $G_{ep}$.

Our numerical method starts with taking $k_z$ for $n_z = 1$ in Eq. (S4) as a guess set of $K_z$ to evaluate $k_\mu$ by solving Eq. (S8) in the main manuscript. We then compare $k_z$ with $k_\mu$; if $k_z < k_\mu$, then we increase the size of $K_z$ by increasing $n_z$ one more. The iteration repeats till the largest $k_z$ in $K_z$ is more than calculated $k_\mu$. The results for corresponding chemical potential μ (or temperature-dependent Fermi energy) and DOS at Fermi energy are plotted in figure S2. Both μ and $g_F$ are dimensionless, normalized with their corresponding bulk values. Generalizing the methods in Ref. [7] to this temperature-dependent Fermi energy case, we found how the $L_z$ dependence of DOS at Fermi level changes with temperature, to be able to take into account this size effect for $G_{ep}$.

In figure S2, the chemical potential change with electron temperature, and Fermi level density of state, are shown. In addition, the electron phonon coupling coefficient is shown in figure S2(c). Both DOS$_F$ and $G_{ep}$ are independent of the electron temperature at the temperature ranges applicable in our calculations (where maximum electron temperature does not exceed 2000 K).

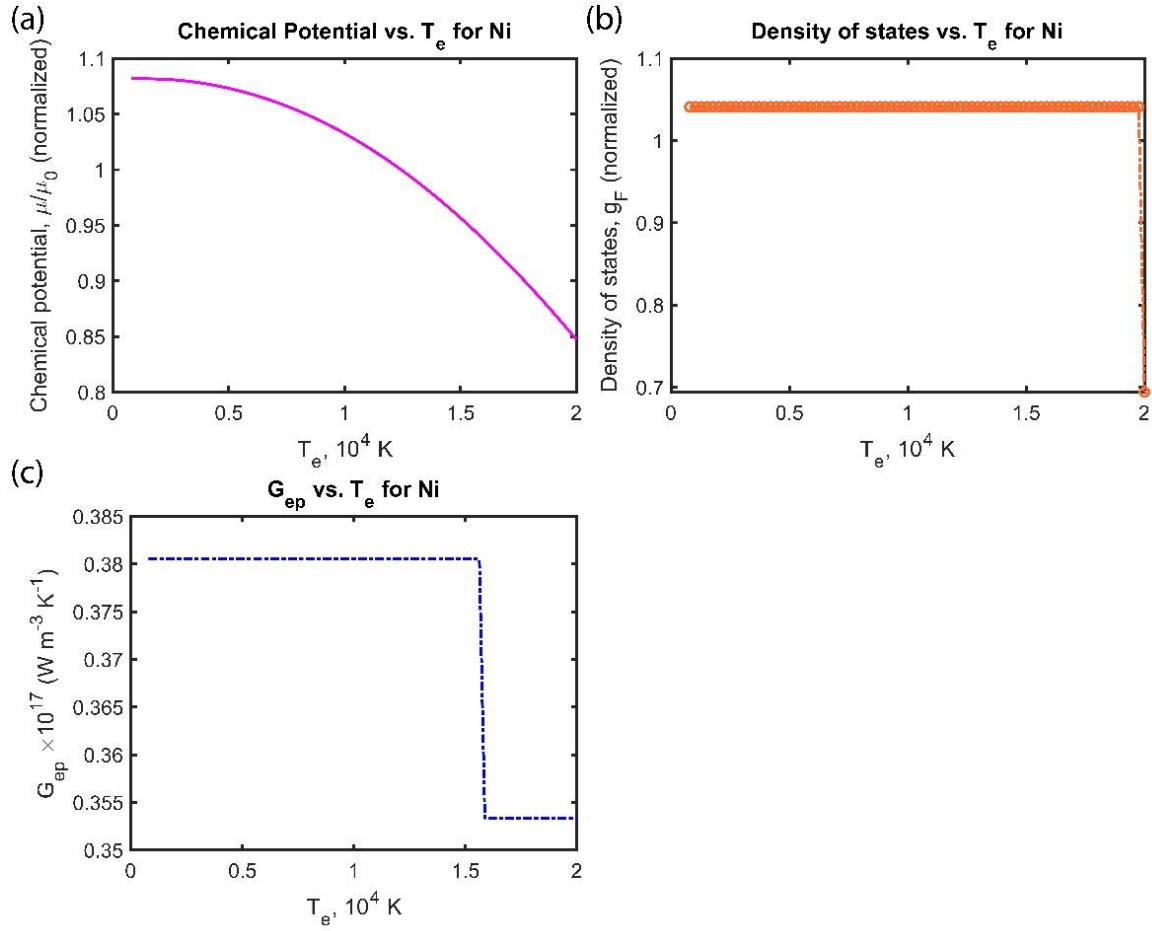

**Figure S2**. The dependence of the Fermi level (**a**) chemical potential (m), (**b**) density of states, and (**c**) electron phonon coupling ($G_{ep}$) on the electron temperature ($T_e$).

The effect of substrate (especially if it is non-magnetic) might be negligible in changing of the magnetization order. However, it might influence the contributing electron and phonon density and so $G_{ep}$. Nevertheless, assuming an isolated thin film is reasonable considering the previous studies [8,9].

In figure 4 of the main manuscript, we showed the magnetization dynamics of the Ni ultrathin film for different film thicknesses. A zoomed in version of the magnetization dynamics at a smaller time ranges are shown in figure S3.

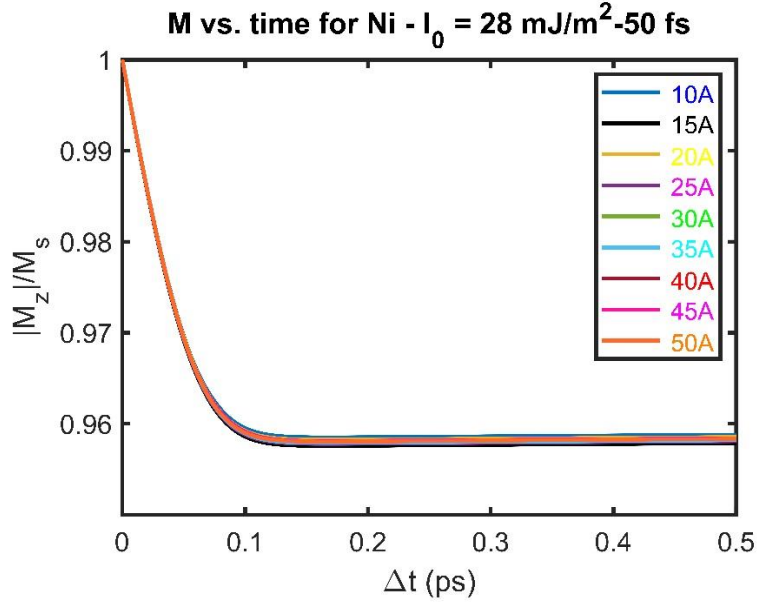

**Figure S3**. (Color online) Magnetization dynamics of the Ni ultrathin film in 300 fs after interaction with the laser pulse.

## Sensitivity of quantum confinement effect of magnetic ultrathin films to the potential well depth ($V_z$)

In the main manuscript, we reported our calculation results of quantum confinement effect in figure, 2 and 4, using the potential well depth of $V_z$= 10 eV. In this section of SI document, we present the sensitivity of our results to the size of the potential well. Practically, the $V_z$ is an infinite value, however, in the FEM calculations, a finite value is given to it. We consider three extra cases of $V_z$=5 eV, 15 eV, and 20 eV.

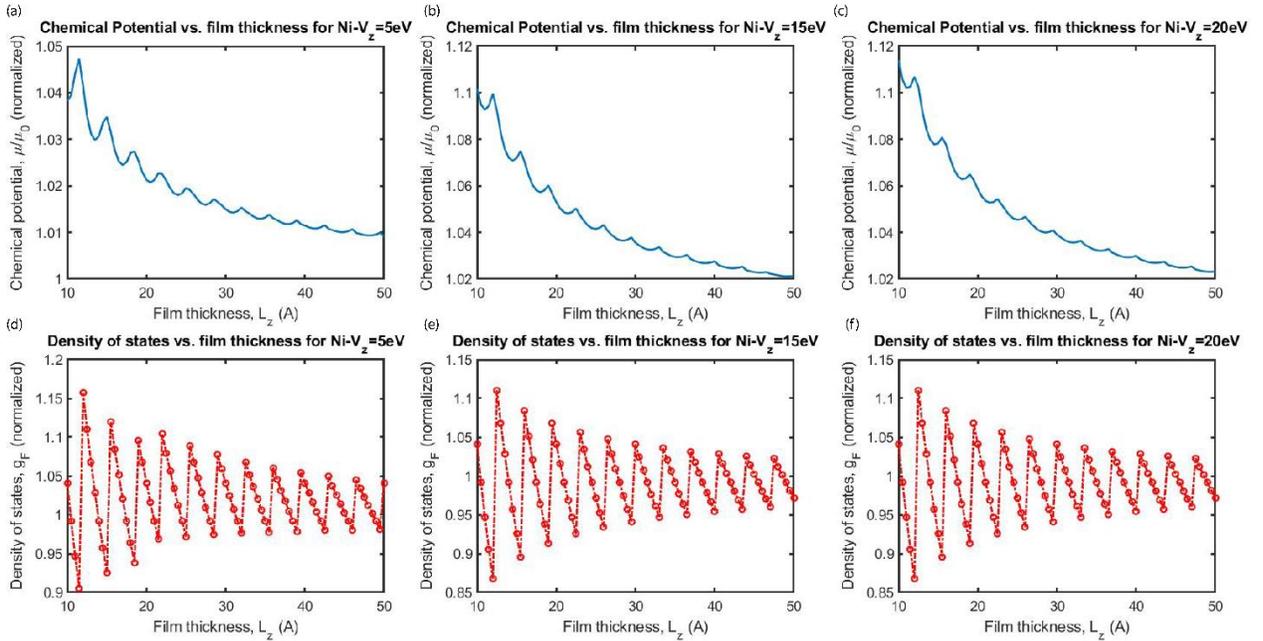

**Figure S4**. Thickness dependent (**a**) chemical potential of the quantum confined Ni ultrathin film with quantum well depth of $V_z$=5 eV, (**b**) 15 eV, (**c**) 20 eV, (**d**) Fermi level electron density of states for $V_z$= 5 eV (**e**) 15 eV (**f**) 20 eV.

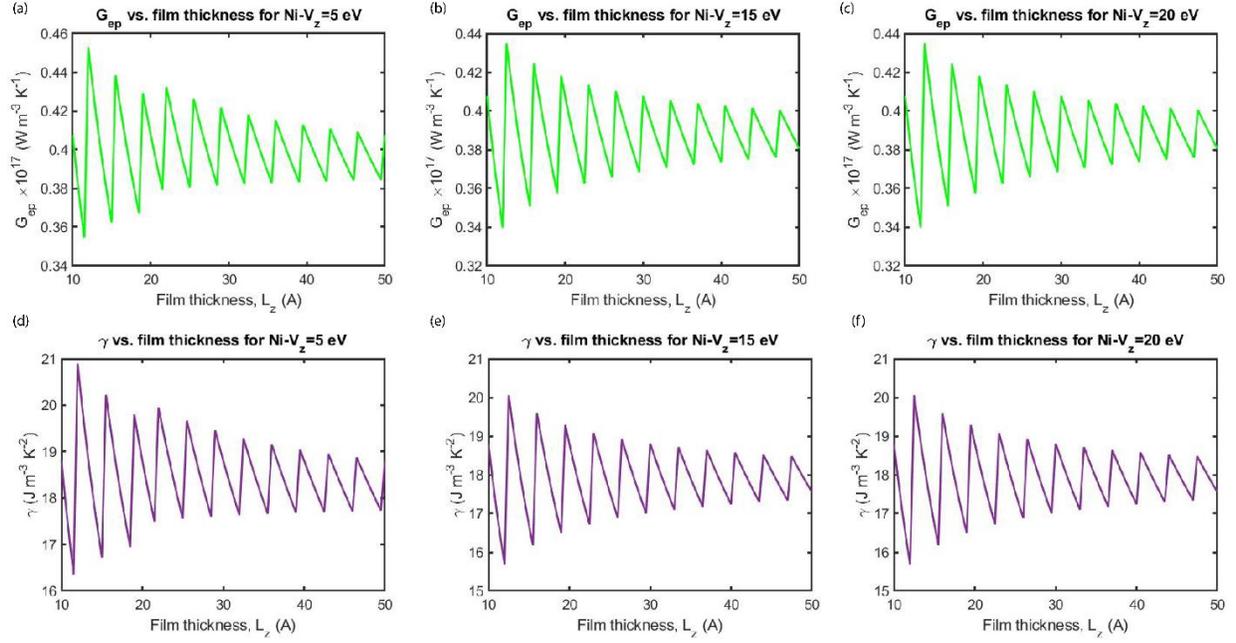

**Figure S5**. Thickness dependent (**a**) electron-phonon coupling coefficient the quantum confined Ni ultrathin film with quantum well depth of $V_z$=5 eV, (**b**) 15 eV, (**c**) 20 eV, (**d**) Sommerfeld coefficient for $V_z$= 5 eV (**e**) 15 eV (**f**) 20 eV.

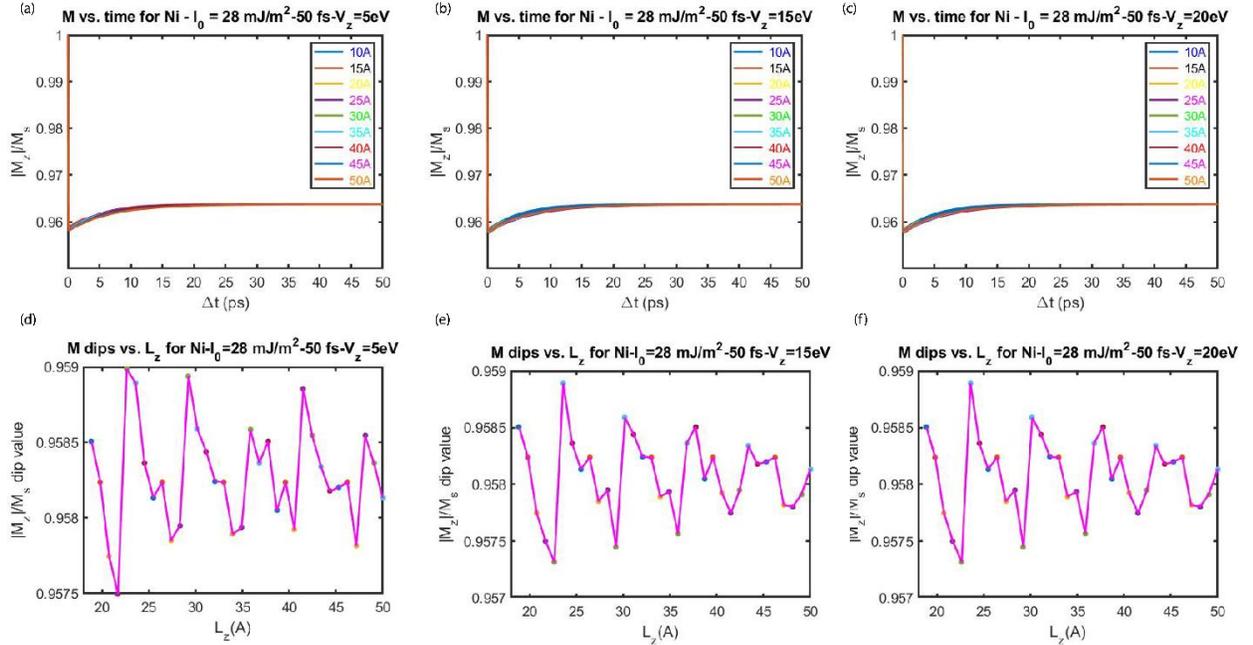

**Figure S6**. Thickness dependent (**a**) magnetization dynamics of the quantum confined Ni ultrathin film with quantum well depth of $V_z$=5 eV, (**b**) 15 eV, (**c**) 20 eV, (**d**) demagnetization dip for $V_z$= 5 eV (**e**) 15 eV (**f**) 20 eV.

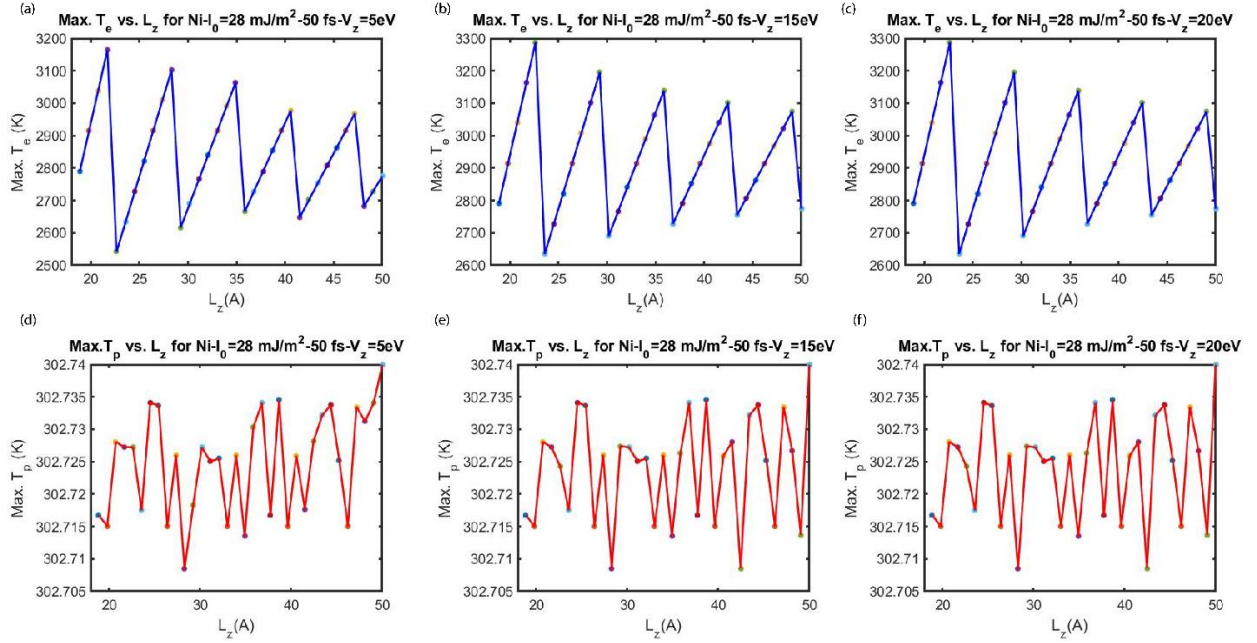

**Figure S7**. Thickness dependent (**a**) maximum electron temperature of the quantum confined Ni ultrathin film with quantum well depth of $V_z=5$ eV, (**b**) 15 eV, (**c**) 20 eV, (**d**) maximum phonon temperature for $V_z=$ 5 eV (**e**) 15 eV (**f**) 20 eV.

The calculations results for the various potential well depths ($V_z=5$, 10, 15, and 20 eV) shows that only in case of sub-20 Å film thicknesses, $V_z$ size might result in a few percent difference in the calculations. However, it does not influence the qualitative predictions of our model on decrease and oscillations in the electron density of states and electron-phonon coupling as the result of quantum confinement effect. In other words μ increases slightly (~ 3%), in sub-20 Å range, from $V_z=5$ eV to $V_z=10$ eV, but the oscillations are not considerable when $V_z$ changes from 10 eV to 15 eV or 20 eV. Accordingly, $g_F$, $G_{ep}$, and γ oscillations decrease (~ 4%) changing from $V_z=5$ eV to $V_z=10$ eV, while remains unchanged for other $V_z$ values. We also investigated the sensitivity of the figure 4 results to the size of $V_z$. Despite similar negligible variations for $V_z=$ 5 eV in case of low film thicknesses, changing the depth of the potential well did not influence the conclusions of the quantum confinement effect on the oscillations of the magnetization dynamics as well as electron and phonon temperatures (figure S6-S7). Increasing the $V_z$ from 5 eV, a small increase (~ 3%) in the maximum electron temperature and decrease in demagnetization dip was observed for sub-20 Å thin films. However, for other $V_z$ values, the results remained unchanged. Since the decrease in the chemical potential and consequently the $g_F$, and $G_{ep}$, is averagely monotonic [7] (the amplitude of the oscillations does not exceed a few percent), the qualitative results of our calculations are consistent using different potential well depths.

**M3TM model considering spin-phonon and spin-electron scattering**

In this section, we calculated the effect of film thickness including the spin-phonon, and spin-electron scattering in the M3TM models shown in figure S8 (a). We used the following equations:

$$C_e \frac{dT_e}{dt} = -G_{ep}(T_e - T_p) - G_{es}(T_e - T_s) + P(t) \qquad (1)$$

$$C_p \frac{dT_p}{dt} = -G_{ep}(T_p - T_e) - G_{ps}(T_p - T_s) \qquad (2)$$

$$C_s \frac{dT_s}{dt} = -G_{es}(T_s - T_e) - G_{ps}(T_s - T_p) \qquad (3)$$

$$\frac{dm}{dt} = Rm\frac{T_p}{T_C}\left(1 - m\coth(m\frac{T_C}{T_e})\right) \qquad (4)$$

where $C_s$ is the spin heat capacity, $G_{es}$, and $G_{ps}$ are the electron-spin and phonon-spin coupling coefficients, respectively. $T_s$ represents the spin temperature. Note that both $G_{ep}$ and R change with the film thickness, as mentioned in the manuscript. if we heuristically assume that $G_{es}$ and $G_{ps}$ have similar size dependence as with $G_{ep}$, through their dependence to $DOS_F$, we find that our results with constant $G_{ep}$, and $G_{es}$ do not change significantly. We therefore intuitively expect that the size dependence of $G_{es}$ and $G_{ps}$ are negligible on $T_e$ and $T_p$ dynamics, which determine the magnetization evolution.

According to the figure S8(a), due to the energy loss to the spin-phonon scattering phenomena, the laser fluence needed to recover the magnetization is larger compared to the condition where the spin-lattice scattering is neglected ($I_0$=35 mJ·m$^{-2}$). In addition, the magnetization drop time is slightly longer, while the recovered magnetization ratio is slightly lower as the result of spin scattering. Figure S8 (b) shows the thickness dependence of the lattice temperature. According to this figure, the maximum lattice temperature increases due to the energy exchange between spin and phonon baths. However, maximum $T_p$ does not exceed 305.976 K, which is still well below the Curie temperature of Ni.

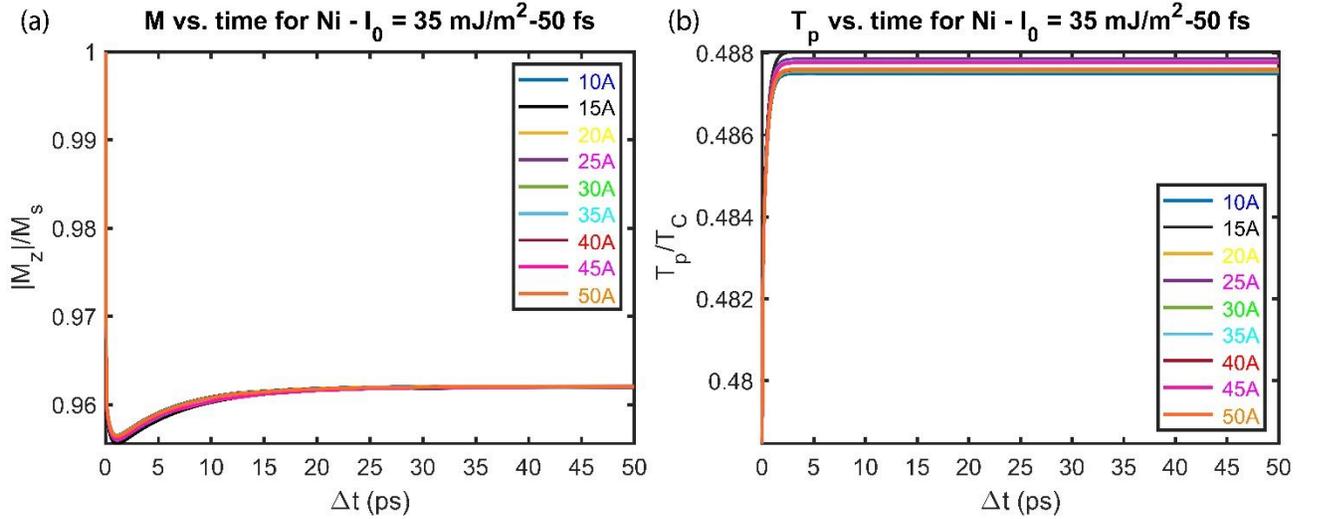

**Figure S8.** Effect of Ni film thickness on (**a**) magnetization dynamics for $L_z$ = 10 Å, 15 Å, 20 Å, 25 Å, 30 Å, 35 Å, 40 Å, 45 Å, and 50 Å thick Ni films, (**b**) $T_p/T_C$, illuminated with $I_0$ = 35 mJ·m-2 Gaussian single laser pulse.

In the previous studies, the easy axis of the Ni changes from perpendicular to in-plane when the film thickness changes from around 12-14 Å to 24 Å [10,11]. The size-dependent change in the magnetic anisotropy is due to shape anisotropy (dominant anisotropy in the thicker films that dictate the in-plane easy axis), which is completely geometry dependent. In our magnetization dynamics model, the "m" stands for the normalized magnetization vector magnitude ($|M_z/M_s|$). After interaction with the low-fluence fs laser pulse, the magnetization vector orientation might change but not switch completely. The motion is described as precession, canting, or maybe toggle from an initial stable orientation, which we assume to be $m_0$=1, regardless of the direction. The magnetization magnitude, which changes due to spin reorientation, does not influence the temperature dynamics in the 2TM or M3TM. The easy axis and magnetic anisotropy effects are studied using LLG equations where the magnetization is a vector, which is beyond our paper's scope [12-15].